\documentclass[12pt]{article}

\usepackage{cite}
\usepackage{graphicx}
\usepackage{pict2e}
\author{Yu.~M.~Zinoviev
       \thanks{E-mail address: Yurii.Zinoviev@ihep.ru} \\[0.5cm]
        {\it Institute for High Energy Physics} \\
        {\it of National Research Center "Kurchatov Institute"} \\
        {\it Protvino, Moscow Region, 142280, Russia}}

\title{On massive higher spins and gravity. III. Spin 7/2}

\date{}

\begin{document}

\maketitle

\begin{abstract}
In this paper, we extend our previous results on the gravitational
interactions for massive spin 5/2 particles and spin 3 particles to
massive spin 7/2, including its massless and partially massless
limits. These results share some common features, such as a 
non-singular massless limit in $AdS$ and a flat limit for non-zero
masses, as well as a singularity at the points corresponding to the
boundary of the unitary forbidden region. At the same time, these
results allow us to suggest what the structure of non-minimal
interactions for arbitrary spins looks like. Another subject of
interest is the Skvortsov-Vasiliev formalism for describing partially
massless fields. This formalism has been very useful in our research,
but our examples have shown that it dos not always lead to the correct
results.
\end{abstract}

\thispagestyle{empty}
\newpage
\setcounter{page}{1}

\section{Introduction}

In the recent two papers \cite{Zin25a,Zin25b} we elaborated on the
gravitational interactions for the massive spin 5/2 and spin 3 fields.
To guarantee a correct number of physical degrees of freedom when
switching on interactions, we use a frame-like gauge invariant
description of massive higher spins \cite{Zin08b,PV10,KhZ19}, which
requires the introduction of Stueckelberg fields, leading to technical
but important problems related to field redefinitions
\cite{BDGT18,Zin24a}. To resolve these ambiguities, we restrict
ourselves to minimal vertices, i.e. vertices that contain the results
from standard substitution rules as well as non-minimal terms with
the lowest number of derivatives possible. To discover the structure of
the necessary non-minimal terms, it was useful to consider
interactions for massless and so-called partially massless cases.
There are no Stueckelberg fields in the massless case, while for the
partially massless cases we use the Skvortsov-Vasiliev formalism 
developed for bosonic fields in \cite{SV06} and extended to the
fermionic fields in \cite{KhZ19}. 

The results obtained for massive spin 5/2 and spin 3 have some common
properties. In anti de Sitter space $\Lambda < 0$ the vertices have
non-singular massless limits, while for non-zero masses they have 
non-singular flat limits. In both cases we found a singularity
corresponding to the points at the boundary of the so-called unitary
forbidden regions in de Sitter space $\Lambda > 0$. The only
way to obtain non-trivial interactions at these points was to
rescale the coupling constant, leaving only non-minimal
terms. It seems reasonable to assume that these properties 
hold for arbitrary spins. However, in both cases considered
only the highest helicities require non-minimal interactions, 
so it's not clear how the structure of these terms looks for
arbitrary spin. Therefore, in this paper we consider another concrete
example: massive spin 7/2 (also including its massless and
partially massless limits).

One more reason why we think it is instructive to consider another
concrete example is related to the Skvortsov-Vasiliev formalism. The
existence of such a formalism is due to the fact that, in the
partially massless case, some gauge symmetries remains unbroken, and
some of the gauge fields don't have their Stueckelberg fields. 
Starting with the general formalism (i.e. formalism obtained from the
general massive case for special value of mass) we make partial gauge
fixing \cite{KhZ19} setting all the remaining Stueckelberg fields to
zero and solving their equations. The resulting formalism doesn't
contain any Stueckelberg fields and works without ambiguities. 
However, to ensure the correct number of physical degrees of
freedom, we must take care on gauge symmetries that were fixed.
Fortunately, for spins 5/2 and 3, all variations
of vertices obtained under these gauge transformations can be
compensated by appropriate corrections to the graviton
transformations. But we cannot be sure that this will always be the
case, and in this paper, we provide an example where it is not.

The organization of the paper is straightforward. In Section 2 we
provide all necessary kinematical information about the gauge
invariant frame-like description for the massive spin 7/2 particles,
including their massless and partially massless limits. We explicitly
show how the Skvortsov-Vasiliev formalism can be derived from the
general one. Then, in Section 3, we consider gravitational
interactions for massless, partially massless and massive cases,
ordering them according to the number of physical degrees of freedom.

\section{Kinematics}

In this section, we provide all the necessary kinematic information on
the gauge invariant frame-like description \cite{Zin08b,PV10,KhZ19} of
massive spin 7/2 including all its massless and partially massless
limits. We work with the multispinor version of the formalism
\cite{KhZ19} and use the same notation and conventions as in the
previous two papers \cite{Zin25a,Zin25b}.

\subsection{Massive case}

Massive spin 7/2 in $d=4$ contains eight  helicities $(\pm 7/2, \pm
5/2, \pm 3/2, \pm 1/2)$ so for its gauge invariant description we need
three one-forms $\Phi^{\alpha(3)\dot\alpha(2)} + h.c$, 
$\Phi^{\alpha(2)\dot\alpha} + h.c.$, $\Phi^\alpha + h..c.$ and 
zero-form $\phi^\alpha + h.c.$. The free Lagrangian has the form:
\begin{eqnarray}
{\cal L}_0 &=& \sum_{k=1}^3 (-1)^k 
\Phi_{\alpha(k-1)\beta\dot\alpha(k-1)} e^\beta{}_{\dot\beta}
D \Phi^{\alpha(k-1)\dot\alpha(k-1)\dot\beta} - \phi_\alpha
E^\alpha{}_{\dot\alpha} D \phi^{\dot\alpha} \nonumber \\
 && + \sum_{k=2}^3 (-1)^k c_k E^{\beta(2)} 
\Phi_{\alpha(k-2)\beta(2)\dot\alpha(k-1)} 
\Phi^{\alpha(k-2)\dot\alpha(k-1)} + c_0 \Phi_\alpha 
E^\alpha{}_{\dot\alpha} \phi^{\dot\alpha} \nonumber \\
 && + \sum_{k=1}^3 (-1)^k \frac{2M}{k(k+1)} [ (k+1)
\Phi_{\alpha(k-1)\beta\dot\alpha(k-1)} E^\beta{}_\gamma
\Phi^{\alpha(k-1)\gamma\dot\alpha(k-1)} \nonumber \\
 && \qquad - (k-1) \Phi_{\alpha(k)\dot\alpha(k-2)\dot\beta}
E^{\dot\beta}{}_{\dot\gamma}
\Phi^{\alpha(k)\dot\alpha(k-2)\dot\gamma}] + 2M E \phi_\alpha
\phi^\alpha + h.c. \label{lag}
\end{eqnarray}
where
\begin{equation}
M^2 = m^2 - 9\Lambda, \qquad
c_3{}^2 = \frac{7}{9}m^2, \qquad
c_2{}^2 = 3(m^2 - 5\Lambda), \qquad
c_0{}^2 = 30(m^2 - 8\Lambda). \label{param}
\end{equation}
This Lagrangian is invariant under the following local gauge
transformations:
\begin{eqnarray}
\delta \Phi^{\alpha(3)\dot\alpha(2)} &=& 
D \zeta^{\alpha(3)\dot\alpha(2)} + e_\beta{}^{\dot\alpha}
\zeta^{\alpha(3)\beta\dot\alpha} + \frac{M}{3} e^\alpha{}_{\dot\beta}
\zeta^{\alpha(2)\dot\alpha(2)\dot\beta} + \frac{c_3}{8}
e^{\alpha\dot\alpha} \zeta^{\alpha(2)\dot\alpha}, \nonumber \\
\delta \Phi^{\alpha(2)\dot\alpha} &=& D \zeta^{\alpha(2)\dot\alpha} +
e_\beta{}^{\dot\alpha} \zeta^{\alpha(2)\beta} + c_3 e_{\beta\dot\beta}
\zeta^{\alpha(2)\beta\dot\alpha\dot\beta} + \frac{2M}{3}
e^\alpha{}_{\dot\beta} \zeta^{\alpha\dot\alpha\dot\beta} +
\frac{c_2}{3} e^{\alpha\dot\alpha} \Phi^\alpha, \nonumber \\
\delta \Phi^\alpha &=& D \zeta^\alpha + c_2 e_{\beta\dot\alpha}
\zeta^{\alpha\beta\dot\alpha} + 2M e^\alpha{}_{\dot\alpha}
\Phi^{\dot\alpha} - \frac{c_0}{3} E^\alpha{}_\beta \phi^\beta,
\label{gauge} \\
\delta \phi^\alpha &=& c_0 \zeta^\alpha. \nonumber 
\end{eqnarray}
But to construct a complete set of the gauge invariant objects
(curvatures) we need the so-called extra fields\footnote{Extra fields
do not enter the free Lagrangian but  on-shell they equivalent to
higher derivatives of physical fields and so play an important role in
the construction of interactions}: one-forms
$\Phi^{\alpha(5)}$, $\Phi^{\alpha(4)\dot\alpha}$, 
$\Phi^{\alpha(3)}$ and Stueckelberg zero-forms $\phi^{\alpha(5)}$,
$\phi^{\alpha(4)\dot\alpha}$, $\phi^{\alpha(3)\dot\alpha(2)}$,
$\phi^{\alpha(3)}$ and $\phi^{\alpha(2)\dot\alpha}$ with the following
gauge transformations:
\begin{eqnarray}
\delta \Phi^{\alpha(5)} &=& D \zeta^{\alpha(5)} + \frac{c_2{}^2}{15}
e^\alpha{}_{\dot\alpha} \zeta^{\alpha(4)\dot\alpha}, \nonumber \\
\delta \Phi^{\alpha(4)\dot\alpha} &=& D \zeta^{\alpha(4)\dot\alpha}
+ e_\beta{}^{\dot\alpha} \zeta^{\alpha(4)\beta} + \frac{c_0{}^2}{240}
e^\alpha{}_{\dot\beta} \zeta^{\alpha(3)\dot\alpha\dot\beta} +
\frac{c_3}{40} e^{\alpha\dot\alpha} \zeta^{\alpha(3)}, \\
\delta \Phi^{\alpha(3)} &=& D \zeta^{\alpha(3)} + 3c_3
e_{\beta\dot\alpha} \zeta^{\alpha(3)\beta\dot\alpha} + 
\frac{c_0{}^2}{48} e^\alpha{}_{\dot\alpha} 
\zeta^{\alpha(2)\dot\alpha}, \nonumber 
\end{eqnarray}
\begin{eqnarray}
\delta \phi^{\alpha(5)} &=& \zeta^{\alpha(5)}, \qquad
\delta \phi^{\alpha(4)\dot\alpha} = \zeta^{\alpha(4)\dot\alpha},
\qquad \delta \phi^{\alpha(3)\dot\alpha(2)} = 
\zeta^{\alpha(3)\dot\alpha(2)}, \nonumber \\
\quad \delta \phi^{\alpha(3)} &=& \zeta^{\alpha(3)}, \qquad
\delta \phi^{\alpha(2)\dot\alpha} = \zeta^{\alpha(2)\dot\alpha}.
\end{eqnarray}
Note that in the complete set of fields we have a one-to-one
correspondence between one-form gauge fields and zero-form
Stueckelberg fields (see Figure 1) in agreement with the fact that in
the massive case all gauge symmetries are spontaneously broken.
\begin{figure}[htb]
\begin{center}
\begin{picture}(180,80)

\put(10,10){\vector(1,0){60}}
\put(10,10){\vector(0,1){60}}
\put(0,60){\makebox(10,10)[]{$N$}}
\put(60,0){\makebox(10,10)[]{$\dot{N}$}}

\put(10,60){\circle*{2}}
\put(10,60){\circle{4}}

\multiput(10,40)(10,10){2}{\circle*{2}}
\multiput(10,40)(10,10){2}{\circle{4}}

\multiput(10,20)(10,10){3}{\circle*{2}}
\multiput(10,20)(10,10){3}{\circle{4}}

\multiput(20,10)(10,10){3}{\circle*{2}}
\multiput(20,10)(10,10){3}{\circle{4}}

\multiput(40,10)(10,10){2}{\circle*{2}}
\multiput(40,10)(10,10){2}{\circle{4}}

\put(60,10){\circle*{2}}
\put(60,10){\circle{4}}

%%%%%%%%%%%%%%%%%%%%%%%%%%%%%%%%%%%%%%%%%%%%%%%%%%%%%%%%

\put(100,30){\vector(1,0){70}}
\put(100,10){\vector(0,1){60}}
\put(90,55){\makebox(10,10)[]{$\Lambda$}}
\put(170,25){\makebox(10,10)[]{$m^2$}}

\put(100,30){\line(3,1){70}}
\put(100,30){\line(2,1){60}}
\put(100,30){\line(1,1){30}}
\put(128,60){\makebox(10,10)[]{$m^2=5\Lambda$}}
\put(158,60){\makebox(10,10)[]{$m^2=8\Lambda$}}
\put(165,52){\makebox(10,10)[]{$m^2=9\Lambda$}}

\end{picture}
\caption{Massive spin 7/2. 1a) Spectrum of fields: $\dot{N},N$ are
numbers of dotted and undotted spinor indices, dots stand for gauge
fields, circles -- for Stueckelberg fields. 1b) Unitary forbidden
region in $dS$ and two partially massless cases.}
\end{center}
\end{figure}
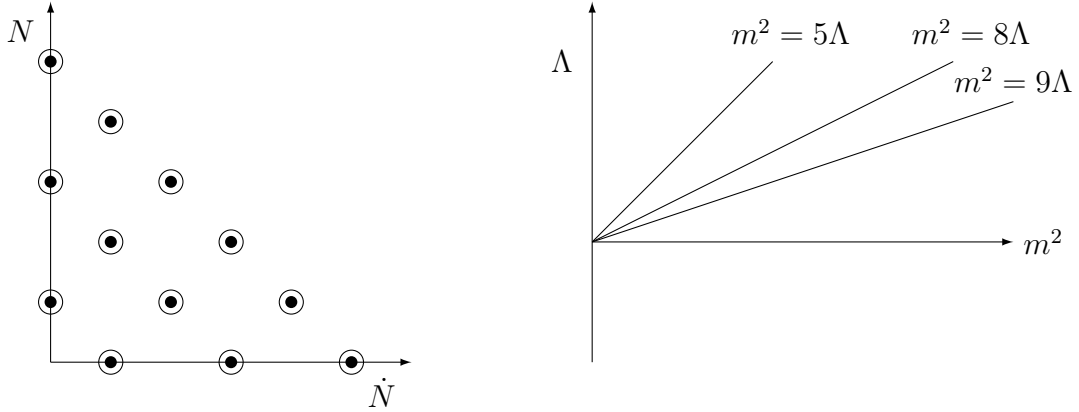
The gauge invariant two-forms for the gauge fields have the following
form:
\begin{eqnarray}
{\cal F}^{\alpha(5)} &=& D \Phi^{\alpha(5)} + \frac{c_2{}^2}{15}
e^\alpha{}_{\dot\alpha} \Phi^{\alpha(4)\dot\alpha} - \frac{2m^2}{5}
E^\alpha{}_\beta \phi^{\alpha(4)\beta} - \frac{c_3c_2{}^2}{50}
E^{\alpha(2)} \phi^{\alpha(3)}, \nonumber \\
{\cal F}^{\alpha(4)\dot\alpha} &=& D \Phi^{\alpha(4)\dot\alpha} +
e_\beta{}^{\dot\alpha} \Phi^{\alpha(4)\beta} + \frac{c_0{}^2}{240}
e^\alpha{}_{\dot\beta} \Phi^{\alpha(3)\dot\alpha\dot\beta} +
\frac{3c_3}{40} e^{\alpha\dot\alpha} \Phi^{\alpha(3)}, \nonumber \\
{\cal F}^{\alpha(3)\dot\alpha(2)} &=& D \Phi^{\alpha(3)\dot\alpha(2)}
+ e_\beta{}^{\dot\alpha} \Phi^{\alpha(3)\beta\dot\alpha} + \frac{M}{3}
e^\alpha{}_{\dot\beta} \Phi^{\alpha(2)\dot\alpha(2)\dot\beta} +
\frac{c_3}{8} e^{\alpha\dot\alpha} \Phi^{\alpha(2)\dot\alpha},
\label{cur} \\
{\cal F}^{\alpha(3)} &=& D \Phi^{\alpha(3)} + 3c_3 e_{\beta\dot\alpha}
\Phi^{\alpha(3)\beta\dot\alpha} + \frac{c_0{}^2}{48} 
e^\alpha{}_{\dot\alpha}\Phi^{\alpha(2)\dot\alpha} \nonumber \\
 && - \frac{8c_2{}^2}{15} E^\alpha{}_\beta  \phi^{\alpha(2)\beta} -
6c_3 E_{\beta(2)} \phi^{\alpha(3)\beta(2)} - \frac{c_3c_0{}^2}{36}
 E^{\alpha(2)} \phi^\alpha, \nonumber \\
{\cal F}^{\alpha(2)\dot\alpha} &=& D \Phi^{\alpha(2)\dot\alpha} +
e_\beta{}^{\dot\alpha} \Phi^{\alpha(2)\beta} + c_3
e_{\beta\dot\beta} \Phi^{\alpha(2)\beta\dot\alpha\dot\beta} +
\frac{2M}{3} e^\alpha{}_{\dot\beta} \Phi^{\alpha\dot\alpha\dot\beta} +
\frac{c_2}{3} e^{\alpha\dot\alpha} \Phi^\alpha, \nonumber \\
{\cal F}^\alpha &=& D \Phi^\alpha + c_2 e_{\beta\dot\alpha}
\Phi^{\alpha\beta\dot\alpha} + 2M e^\alpha{}_{\dot\alpha}
\Phi^{\dot\alpha} - \frac{c_0}{3} E^\alpha{}_\beta \phi^\beta,
\nonumber 
\end{eqnarray}
while the gauge invariant one-forms for the Stueckelberg fields look
as follows
\begin{eqnarray}
{\cal C}^{\alpha(5)} &=& D \phi^{\alpha(5)} - \Phi^{\alpha(5)} 
+ \frac{c_2{}^2}{15} e^\alpha{}_{\dot\alpha} 
\phi^{\alpha(4)\dot\alpha}, \nonumber \\
{\cal C}^{\alpha(4)\dot\alpha} &=& D \phi^{\alpha(4)\dot\alpha}
- \Phi^{\alpha(4)\dot\alpha} + e_\beta{}^{\dot\alpha}
\phi^{\alpha(4)\beta} + \frac{c_0{}^2}{240} e^\alpha{}_{\dot\beta}
\phi^{\alpha(3)\dot\alpha\dot\beta} + \frac{c_3}{40}
e^{\alpha\dot\alpha} \phi^{\alpha(3)}, \nonumber \\
{\cal C}^{\alpha(3)\dot\alpha(2)} &=& D \phi^{\alpha(3)\dot\alpha(2)}
- \Phi^{\alpha(3)\dot\alpha(2)} + e_\beta{}^{\dot\alpha} 
\phi^{\alpha(3)\beta\dot\alpha} + \frac{M}{3} e^\alpha{}_{\dot\beta}
\phi^{\alpha(2)\dot\alpha(2)\dot\beta} + \frac{c_3}{8}
e^{\alpha\dot\alpha} \phi^{\alpha(2)\dot\alpha}, \nonumber \\
{\cal C}^{\alpha(3)} &=& D \phi^{\alpha(3)} - \Phi^{\alpha(3)} + 3c_3
e_{\beta\dot\alpha} \phi^{\alpha(3)\beta\dot\alpha} + 
\frac{c_0{}^2}{48} e^\alpha{}_{\dot\alpha} \phi^{\alpha(2)\dot\alpha},
 \\
{\cal C}^{\alpha(2)\dot\alpha} &=& D \phi^{\alpha(2)\dot\alpha} -
\Phi^{\alpha(2)\dot\alpha} + e_\beta{}^{\dot\alpha}
\phi^{\alpha(2)\beta} + c_3 e_{\beta\dot\beta}
\phi^{\alpha(2)\beta\dot\alpha\dot\beta} + \frac{2M}{3}
e^\alpha{}_{\dot\beta} \phi^{\alpha\dot\alpha\dot\beta} +
\frac{c_2}{3} e^{\alpha\dot\alpha} \phi^\alpha, \nonumber \\
{\cal C}^\alpha &=& D \phi^\alpha - c_0 \Phi_\alpha  + 2M
e^\alpha{}_{\dot\alpha} \phi^{\dot\alpha}. \nonumber 
\end{eqnarray}
On-shell all these curvatures vanish except
\begin{eqnarray}
{\cal F}^{\alpha(5)} &\approx& E_{\beta(2)} Y^{\alpha(5)\beta(2)},
\qquad {\cal C}^{\alpha(5)} \approx e_{\beta\dot\alpha}
Y^{\alpha(5)\beta\dot\alpha}, \nonumber \\
{\cal C}^{\alpha(4)\dot\alpha} &\approx& e_{\beta\dot\beta}
Y^{\alpha(4)\beta\dot\alpha\dot\beta}, \qquad 
{\cal C}^{\alpha(3)\dot\alpha(2)} \approx e_{\beta\dot\beta}
Y^{\alpha(3)\beta\dot\alpha(2)\dot\beta}
\end{eqnarray}
Here fields $Y$ are gauge invariant zero-forms (analogues of the
Weyl tensor in gravity) which are just the first representatives of
four infinite chains of gauge invariant zero-forms satisfying the
so-called unfolded equations \cite{PV10,KhZ19}.

From the relations in (\ref{param}) it follows that in the de Sitter
space $\Lambda > 0$ there exists a so-called unitary forbidden region
$m^2 < 9\Lambda$ (see Figure 1). Inside this forbidden region there
are two partially massless cases: $m^2 = 8\Lambda$ with helicities
$(\pm 7/2, \pm 5/2, \pm 3/2)$ and $m^2 = 5\Lambda$ with $(\pm 7/2, \pm
5/2)$ only. 

\subsection{First partially massless case}

It corresponds to $m^2 = 8\Lambda$ hence $c_0 = 0$. In the Lagrangian
the zero-form $\phi^\alpha$ completely decouples leaving only the
helicities $(\pm 7/2, \pm 5/2, \pm 3/2)$. At the same time, an
analysis of the gauge invariant curvatures shows that Stueckelberg
zero-forms $\phi^{\alpha(3)\dot\alpha(2)}$ and
$\phi^{\alpha(2)\dot\alpha}$ also decouple though the remaining
curvatures are still gauge invariant. Thus in this case some of the
gauge one-forms do not have their corresponding Stueckelberg
zero-forms and some of the gauge symmetries remain unbroken (see
Figure 2).
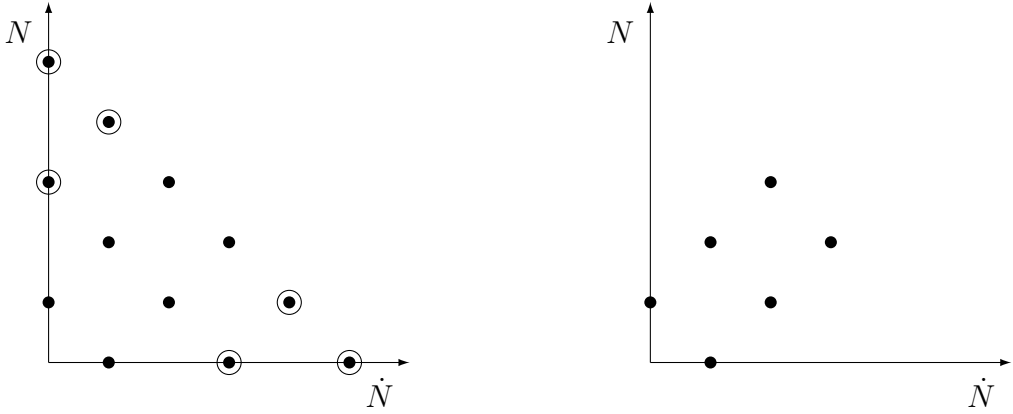
\begin{figure}[htb]
\begin{center}
\begin{picture}(180,80)

\put(10,10){\vector(1,0){60}}
\put(10,10){\vector(0,1){60}}
\put(0,60){\makebox(10,10)[]{$N$}}
\put(60,0){\makebox(10,10)[]{$\dot{N}$}}

\put(10,60){\circle*{2}}
\put(10,60){\circle{4}}

\multiput(10,40)(10,10){2}{\circle*{2}}
\multiput(10,40)(10,10){2}{\circle{4}}

\multiput(10,20)(10,10){3}{\circle*{2}}

\multiput(20,10)(10,10){3}{\circle*{2}}

\multiput(40,10)(10,10){2}{\circle*{2}}
\multiput(40,10)(10,10){2}{\circle{4}}

\put(60,10){\circle*{2}}
\put(60,10){\circle{4}}

%%%%%%%%%%%%%%%%%%%%%%%%%%%%%%%%%%%%%%%%%%%%%%%%%%%%%%%%

\put(110,10){\vector(1,0){60}}
\put(110,10){\vector(0,1){60}}
\put(100,60){\makebox(10,10)[]{$N$}}
\put(160,0){\makebox(10,10)[]{$\dot{N}$}}

\multiput(110,20)(10,10){3}{\circle*{2}}

\multiput(120,10)(10,10){3}{\circle*{2}}

\end{picture}
\caption{First partially massless case. 2a) Before gauge fixing. 2b)
After gauge fixing.}
\end{center}
\end{figure}
Now, to constrict a spin 7/2 analogue of the Skvortsov-Vasiliev
formalism, let us consider partial gauge fixing. It means that we set
Stueckelberg zero-forms $\phi^{\alpha(5)}$,
$\phi^{\alpha(4)\dot\alpha}$ and $\phi^{\alpha(3)}$ equal to zero and
solve their equations:
\begin{eqnarray}
0 &\approx& D \phi^{\alpha(5)} - \Phi^{\alpha(5)} + \frac{c_2{}^2}{15}
e^\alpha{}_{\dot\alpha} \phi^{\alpha(4)\dot\alpha} - 
e_{\beta\dot\alpha} Y^{\alpha(5)\beta\dot\alpha}, \nonumber \\
0 &\approx& D \phi^{\alpha(4)\dot\alpha} - \Phi^{\alpha(4)\dot\alpha}
 + e_\beta{}^{\dot\alpha} \phi^{\alpha(4)\beta} + \frac{c_3}{40}
e^{\alpha\dot\alpha} \phi^{\alpha(3)} - e_{\beta\dot\beta}
Y^{\alpha(4)\beta\dot\alpha\dot\beta}, \\
0 &\approx& D \phi^{\alpha(3)} - \Phi^{\alpha(3)} + 3c_3
e_{\beta\dot\alpha} \phi^{\alpha(3)\beta\dot\alpha}. \nonumber 
\end{eqnarray}
The remaining curvatures
\begin{eqnarray}
{\cal F}^{\alpha(3)\dot\alpha(2)} &=& D \Phi^{\alpha(3)\dot\alpha(2)}
 + \frac{M}{3} e^\alpha{}_{\dot\beta} 
\Phi^{\alpha(2)\dot\alpha(2)\dot\beta} + \frac{c_3}{8}
e^{\alpha\dot\alpha} \Phi^{\alpha(2)\dot\alpha}, \nonumber \\
{\cal F}^{\alpha(2)\dot\alpha} &=& D \Phi^{\alpha(2)\dot\alpha} +
 c_3 e_{\beta\dot\beta} \Phi^{\alpha(2)\beta\dot\alpha\dot\beta} +
\frac{2M}{3} e^\alpha{}_{\dot\beta} \Phi^{\alpha\dot\alpha\dot\beta} +
\frac{c_2}{3} e^{\alpha\dot\alpha} \Phi^\alpha,  \\
{\cal F}^\alpha &=& D \Phi^\alpha + c_2 e_{\beta\dot\alpha}
\Phi^{\alpha\beta\dot\alpha} + 2M e^\alpha{}_{\dot\alpha}
\Phi^{\dot\alpha} \nonumber 
\end{eqnarray}
are still invariant under the $\zeta^{\alpha(3)\dot\alpha(2)}$,
$\zeta^{\alpha(2)\dot\alpha}$ and $\zeta^\alpha$ transformations, but
not invariant under $\zeta^{\alpha(4)\dot\alpha}$ and
$\zeta^{\alpha(3)}$ ones. However, Lagrangian equations are
proportional to the combinations
$$
{\cal F}_{\alpha(2)\dot\alpha(2)\dot\beta} e_\alpha{}^{\dot\beta},
\qquad {\cal F}_{\alpha\dot\alpha\dot\beta} e_\alpha{}^{\dot\beta},
\qquad {\cal F}_{\dot\alpha} e_\alpha{}^{\dot\alpha},
$$
which are invariant as it should be because the Lagrangian is
invariant. For what follows it is important that now the curvature
${\cal F}^{\alpha(3)\dot\alpha(2)}$ does not vanish on-shell
$$
{\cal F}^{\alpha(3)\dot\alpha(2)} \approx E_{\beta(2)}
Y^{\alpha(3)\beta(2)\dot\alpha(2)}.
$$
One more important fact is that the Lagrangian can be written in the
explicitly gauge invariant form:
\begin{equation}
{\cal L}_0 = \frac{1}{2M} {\cal F}_{\alpha(3)\dot\alpha(2)}
{\cal F}^{\alpha(3)\dot\alpha(2)} - \frac{3}{8M} 
{\cal F}_{\alpha(2)\dot\alpha} {\cal F}^{\alpha(2)\dot\alpha} 
+ \frac{1}{4M} {\cal F}_\alpha {\cal F}^\alpha + h.c.
\end{equation}

\subsection{Second partially massless case}

It corresponds to $m^2 = 5\Lambda$, hence $c_2 = 0$. In the Lagrangian
the one-form $\Phi^\alpha$ and zero-form $\phi^\alpha$ decouple
leaving us with the helicities $(\pm 7/2, \pm 5/2)$. At the same time,
an analysis of the gauge invariant curvatures shows that all
Stueckelberg zero-form except $\phi^{\alpha(5)}$ decouple 
(see Figure 3).
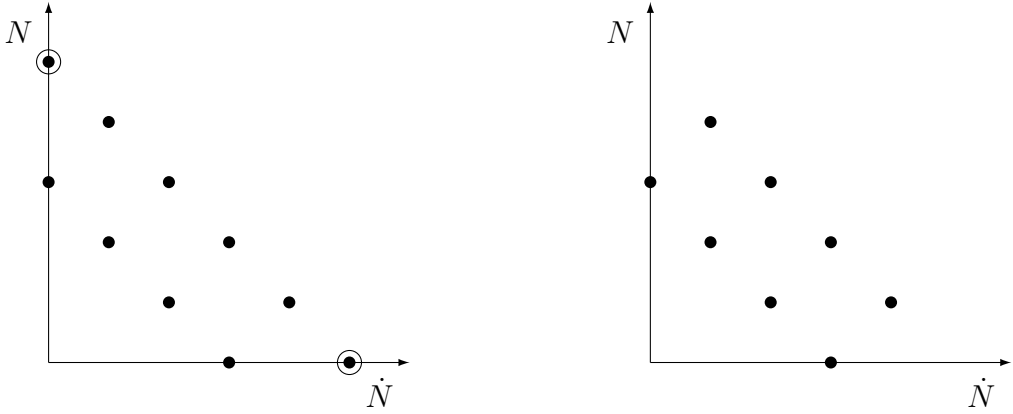
\begin{figure}[htb]
\begin{center}
\begin{picture}(180,80)

\put(10,10){\vector(1,0){60}}
\put(10,10){\vector(0,1){60}}
\put(0,60){\makebox(10,10)[]{$N$}}
\put(60,0){\makebox(10,10)[]{$\dot{N}$}}

\put(10,60){\circle*{2}}
\put(10,60){\circle{4}}

\multiput(10,40)(10,10){2}{\circle*{2}}

\multiput(20,30)(10,10){2}{\circle*{2}}

\multiput(30,20)(10,10){2}{\circle*{2}}

\multiput(40,10)(10,10){2}{\circle*{2}}

\put(60,10){\circle*{2}}
\put(60,10){\circle{4}}

%%%%%%%%%%%%%%%%%%%%%%%%%%%%%%%%%%%%%%%%%%%%%%%%%%%%%%%%

\put(110,10){\vector(1,0){60}}
\put(110,10){\vector(0,1){60}}
\put(100,60){\makebox(10,10)[]{$N$}}
\put(160,0){\makebox(10,10)[]{$\dot{N}$}}

\multiput(110,40)(10,10){2}{\circle*{2}}

\multiput(120,30)(10,10){2}{\circle*{2}}

\multiput(130,20)(10,10){2}{\circle*{2}}

\multiput(140,10)(10,10){2}{\circle*{2}}

\end{picture}
\caption{Second partially massless case. 3a) Before gauge fixing. 3b)
After gauge fixing.}
\end{center}
\end{figure}
Now to reproduce Skvortsov-Vasiliev formalism we set 
$\phi^{\alpha(5)}= 0$ and solve its equation
\begin{equation}
0 \approx D \phi^{\alpha(5)} - \Phi^{\alpha(5)} - e_{\beta\dot\alpha}
Y^{\alpha(5)\beta\dot\alpha}. 
\end{equation}
In this, the remaining curvatures
\begin{eqnarray}
{\cal F}^{\alpha(4)\dot\alpha} &=& D \Phi^{\alpha(4)\dot\alpha} +
 \frac{c_0{}^2}{240} e^\alpha{}_{\dot\beta}
\Phi^{\alpha(3)\dot\alpha\dot\beta} + \frac{3c_3}{40}
e^{\alpha\dot\alpha} \Phi^{\alpha(3)}, \nonumber \\
{\cal F}^{\alpha(3)\dot\alpha(2)} &=& D \Phi^{\alpha(3)\dot\alpha(2)}
+ e_\beta{}^{\dot\alpha} \Phi^{\alpha(3)\beta\dot\alpha} + \frac{M}{3}
e^\alpha{}_{\dot\beta} \Phi^{\alpha(2)\dot\alpha(2)\dot\beta} +
\frac{c_3}{8} e^{\alpha\dot\alpha} \Phi^{\alpha(2)\dot\alpha},
\nonumber \\
{\cal F}^{\alpha(3)} &=& D \Phi^{\alpha(3)} + 3c_3 e_{\beta\dot\alpha}
\Phi^{\alpha(3)\beta\dot\alpha} + \frac{c_0{}^2}{48} 
e^\alpha{}_{\dot\alpha}\Phi^{\alpha(2)\dot\alpha},  \\
{\cal F}^{\alpha(2)\dot\alpha} &=& D \Phi^{\alpha(2)\dot\alpha} +
e_\beta{}^{\dot\alpha} \Phi^{\alpha(2)\beta} + c_3
e_{\beta\dot\beta} \Phi^{\alpha(2)\beta\dot\alpha\dot\beta} +
\frac{2M}{3} e^\alpha{}_{\dot\beta} \Phi^{\alpha\dot\alpha\dot\beta}, 
 \nonumber 
\end{eqnarray}
are invariant under all four gauge transformations
$\zeta^{\alpha(4)\dot\alpha}$, $\zeta^{\alpha(3)\dot\alpha(2)}$,
$\zeta^{\alpha(3)}$ and $\zeta^{\alpha(2)\dot\alpha}$ (but not under
$\zeta^{\alpha(5)}$). Note that in this case the curvature 
${\cal F}^{\alpha(4)\dot\alpha}$ does not vanish on-shell
\begin{equation}
{\cal F}^{\alpha(4)\dot\alpha} \approx E_{\beta(2)}
Y^{\alpha(4)\beta(2)\dot\alpha}. 
\end{equation}
As in the previous case, the Lagrangian can be written as follows
\begin{eqnarray}
{\cal L}_0 &=& - \frac{2}{3M\Lambda} {\cal F}_{\alpha(4)\dot\alpha}
{\cal F}^{\alpha(4)\dot\alpha} + \frac{1}{2M} 
{\cal F}_{\alpha(3)\dot\alpha(2)} {\cal F}^{\alpha(3)\dot\alpha(2)}
\nonumber \\
 && + \frac{1}{15M\Lambda} {\cal F}_{\alpha(3)} {\cal F}^{\alpha(3)}
- \frac{3}{8M} {\cal F}_{\alpha(2)\dot\alpha} 
{\cal F}^{\alpha(2)\dot\alpha} + h.c. 
\end{eqnarray}
where coefficients are chosen so that to satisfy the so-called extra
field decoupling conditions.

\subsection{Massless case}

In this case there are no any Stueckelberg zero-forms and just three
one-forms: $\Phi^{\alpha(5)}$, $\Phi^{\alpha(4)\dot\alpha}$ and
$\Phi^{\alpha(3)\dot\alpha(2)}$. The gauge invariant curvatures look
like:
\begin{eqnarray}
{\cal F}^{\alpha(5)} &=& D \Phi^{\alpha(5)} - \Lambda 
e^\alpha{}_{\dot\alpha} \Phi^{\alpha(4)\dot\alpha}, \nonumber \\
{\cal F}^{\alpha(4)\dot\alpha} &=& D \Phi^{\alpha(4)\dot\alpha}
+ e_\beta{}^{\dot\alpha} \Phi^{\alpha(4)\beta} - \Lambda
e^\alpha{}_{\dot\beta} \Phi^{\alpha(3)\dot\alpha\dot\beta}, \\
{\cal F}^{\alpha(3)\dot\alpha(2)} &=& D \Phi^{\alpha(3)\dot\alpha(2)}
 + e_\beta{}^{\dot\alpha} \Phi^{\alpha(3)\beta\dot\alpha} +
\frac{M}{3} e^\alpha{}_{\dot\beta} 
\Phi^{\alpha(2)\dot\alpha(2)\dot\beta}, \nonumber
\end{eqnarray}
where $M^2 = - 9\Lambda$. In this case only ${\cal F}^{\alpha(5)}$
does not vanish on-shell:
\begin{equation}
{\cal F}^{\alpha(5)} \approx E_{\beta(2)} Y^{\alpha(5)\beta(2)}.
\end{equation}
The free Lagrangian can be written as follows:
\begin{equation}
{\cal L}_0 = \frac{1}{20M\Lambda^2} {\cal F}_{\alpha(5)} 
{\cal F}^{\alpha(5)} - \frac{1}{4M\Lambda} 
{\cal F}_{\alpha(4)\dot\alpha} {\cal F}^{\alpha(4)\dot\alpha}
+ \frac{1}{2M} {\cal F}_{\alpha(3)\dot\alpha(2)}
{\cal F}^{\alpha(3)\dot\alpha(2)} + h.c. 
\end{equation}
and here also the coefficients are chosen according to extra field
decoupling conditions.

\section{Interaction with gravity} 

In this Section we consider gravitational interaction for massive spin
7/2 including all its massless and partially massless limits. We
begin with the simplest massless case and then proceed increasing a
number of physical degrees of freedom.

\subsection{Massless case}

To construct cubic vertex we use the so-called Fradkin-Vasiliev
formalism \cite{FV87,FV87a,Vas11,KhZ20a,Zin24a}. Recall that the first
step in such construction is to find consistent deformations 
$\Delta {\cal F}$ for all gauge invariant curvatures where consistency
means that the deformed curvatures $\hat{\cal F} = {\cal F} + \Delta
{\cal F}$ transform covariantly $\delta \hat{\cal F} \sim {\cal F}
\zeta$. \\
{\bf Deformations for spin 7/2} Not surprisingly, the unique result
here
\begin{eqnarray}
\Delta {\cal F}^{\alpha(5)} &=& \omega^\alpha{}_\beta 
\Phi^{\alpha(4)\beta} - \Lambda h^\alpha{}_{\dot\alpha}
\Phi^{\alpha(4)\dot\alpha}, \nonumber \\
\Delta {\cal F}^{\alpha(4)\dot\alpha} &=& \omega^\alpha{}_\beta
\Phi^{\alpha(3)\beta\dot\alpha} + \omega^{\dot\alpha}{}_{\dot\beta}
\Phi^{\alpha(4)\dot\beta} + h_\beta{}^{\dot\alpha} 
\Phi^{\alpha(4)\beta} - \Lambda h^\alpha{}_{\dot\beta}
\Phi^{\alpha(3)\dot\alpha\dot\beta}, \\
\Delta {\cal F}^{\alpha(3)\dot\alpha(2)} &=& \omega^\alpha{}_\beta
\Phi^{\alpha(2)\beta\dot\alpha(2)} + \omega^{\dot\alpha}{}_{\dot\beta}
\Phi^{\alpha(3)\dot\alpha\dot\beta} + h_\beta{}^{\dot\alpha}
\Phi^{\alpha(3)\beta\dot\alpha} + \frac{M}{3} h^\alpha{}_{\dot\beta}
\Phi^{\alpha(2)\dot\alpha(2)\dot\beta}, \nonumber
\end{eqnarray}
exactly corresponds to the so-called standard substitution rules:
\begin{equation}
e^{\alpha\dot\alpha} \Rightarrow e^{\alpha\dot\alpha} +
h^{\alpha\dot\alpha}, \qquad D \Rightarrow D + \omega^{\alpha(2)}
L_{\alpha(2)} + \omega^{\dot\alpha(2)} L_{\dot\alpha(2)},
\end{equation}
where $L_{\alpha(2)}$, $L_{\dot\alpha(2)}$ are Lorentz group
generators. \\
{\bf Deformations for graviton} Here the results also turn out to be
unique up to one arbitrary coupling constant which must be fixed
later:
\begin{eqnarray}
\Delta R^{\alpha(2)} &=& a_0 [ \Phi^{\alpha\beta(4)} 
\Phi^\alpha{}_{\beta(4)} - 4\Lambda \Phi^{\alpha\beta(3)\dot\alpha}
\Phi^\alpha{}_{\beta(3)\dot\alpha} + 6\Lambda^2 
\Phi^{\alpha\beta(2)\dot\alpha(2)} 
\Phi^\alpha{}_{\beta(2)\dot\alpha(2)} \nonumber \\
 && \quad + 4\Lambda^2 \Phi^{\alpha\beta\dot\alpha(3)} 
\Phi^\alpha{}_{\beta\dot\alpha(3)} - \Lambda
\Phi^{\alpha\dot\alpha(4)} \Phi^\alpha{}_{\dot\alpha(4)} ], \\
\Delta T^{\alpha\dot\alpha} &=& a_0 [ -  \Phi^{\alpha\beta(4)}
\Phi_{\beta(4)}{}^{\dot\alpha} + 4\Lambda 
\Phi^{\alpha\beta(3)\dot\beta} 
\Phi_{\beta(3)\dot\beta}{}^{\dot\alpha}  
 + 2M\Lambda \Phi^{\alpha\beta(2)\dot\beta(2)}
\Phi_{\beta(2)\dot\beta(2)}{}^{\dot\alpha} + h.c.]. \nonumber
\end{eqnarray}
The second step is to consider a deformed Lagrangian $\hat{\cal L}_0$
(i.e. the sum of the free Lagrangians for spin 7/2 and graviton where
all curvatures are replaced by the deformed ones):
\begin{eqnarray}
\hat{\cal L}_0 &=& \frac{1}{20M\Lambda^2} \hat{\cal F}_{\alpha(5)}
\hat{\cal F}^{\alpha(5)} - \frac{1}{4M\Lambda}
\hat{\cal F}_{\alpha(4)\dot\alpha} \hat{\cal F}^{\alpha(4)\dot\alpha}
\nonumber \\
 && + \frac{1}{2M} \hat{\cal F}_{\alpha(3)\dot\alpha(2)}
\hat{\cal F}^{\alpha(3)\dot\alpha(2)} - \frac{1}{4\Lambda}
\hat{R}_{\alpha(2)} \hat{R}^{\alpha(2)}
\end{eqnarray}
and require it to be gauge invariant. Non vanishing on-shell
variations look like
\begin{equation}
\delta \hat{\cal F}^{\alpha(5)} = R^\alpha{}_\beta
\zeta^{\alpha(4)\beta}, \qquad
\delta \hat{R}^{\alpha(2)} = 2a_0 {\cal F}^{\alpha\beta(4)}
\zeta^\alpha{}_{\beta(4)}. 
\end{equation}
They produce
\begin{equation}
\delta \hat{\cal L}_0 = [\frac{1}{2M\Lambda^2} + \frac{2a_0}{\Lambda}]
{\cal F}_{\alpha\gamma(4)} R^\alpha{}_\beta \zeta^{\beta\gamma(4)},
\end{equation}
so we put
$$
a_0 = - \frac{1}{4M\Lambda}.
$$
At last, we extract the cubic part of the deformed Lagrangian
\begin{eqnarray}
M{\cal L}_1 &=& \frac{1}{10\Lambda^2} {\cal F}_{\alpha(5)} 
[ \omega^\alpha{}_\beta \Phi^{\alpha(4)\beta} - \Lambda 
h^\alpha{}_{\dot\alpha} \Phi^{\alpha(4)\dot\alpha} ] \nonumber \\
 && - \frac{1}{2\Lambda}{\cal F}_{\alpha(4)\dot\alpha} 
[ \omega^\alpha{}_\beta \Phi^{\alpha(3)\beta\dot\alpha} +
\omega^{\dot\alpha}{}_{\dot\beta} \Phi^{\alpha(4)\dot\beta} +
h_\beta{}^{\dot\alpha} \Phi^{\alpha(4)\beta} - \Lambda 
h^\alpha{}_{\dot\beta} \Phi^{\alpha(3)\dot\alpha\dot\beta} ] \nonumber
\\
 && + {\cal F}_{\alpha(3)\dot\alpha(2)}  [ \omega^\alpha{}_\beta
\Phi^{\alpha(2)\beta\dot\alpha(2)} + \omega^{\dot\alpha}{}_{\dot\beta}
\Phi^{\alpha(3)\dot\alpha\dot\beta} + h_\beta{}^{\dot\alpha}
\Phi^{\alpha(3)\beta\dot\alpha} + \frac{M}{3} h^\alpha{}_{\dot\beta}
\Phi^{\alpha(2)\dot\alpha(2)\dot\beta} ] \nonumber \\
 && + \frac{1}{8\Lambda^2} R_{\alpha(2)} [ \Phi^{\alpha\beta(4)} 
\Phi^\alpha{}_{\beta(4)} - 4\Lambda \Phi^{\alpha\beta(3)\dot\alpha}
\Phi^\alpha{}_{\beta(3)\dot\alpha} + 6\Lambda^2 
\Phi^{\alpha\beta(2)\dot\alpha(2)} 
\Phi^\alpha{}_{\beta(2)\dot\alpha(2)} \nonumber \\
 && \qquad \qquad + 4\Lambda^2 \Phi^{\alpha\beta\dot\alpha(3)} 
\Phi^\alpha{}_{\beta\dot\alpha(3)} - \Lambda
\Phi^{\alpha\dot\alpha(4)} \Phi^\alpha{}_{\dot\alpha(4)} ].
\end{eqnarray}
Note that formally this vertex contains terms with up to six
derivatives but all such terms (as it common for the massless vertices
\cite{Vas11,KhZ20a}) form total derivatives and can be dropped.
Using explicit expressions for the gauge invariant curvatures,
integrating by parts and using torsion zero condition this vertex can
be transformed into the following on-shell equivalent form:
\begin{eqnarray}
{\cal L}_1 &=& R_{\alpha\beta} [ - \frac{1}{4\Lambda} 
\Phi^{\alpha\dot\alpha(4)} \Phi^\beta{}_{\dot\alpha(4)} +
\Phi^{\alpha\gamma\dot\alpha(3)} \Phi^\beta{}_{\gamma\dot\alpha(3)}]
\nonumber \\
 && + D \Phi_{\alpha\beta(2)\dot\alpha(2)} h^\alpha{}_{\dot\beta}
\Phi^{\beta(2)\dot\alpha(2)\dot\beta} 
 + \frac{1}{3} \Phi_{\alpha(2)\dot\alpha(2)\dot\beta}
e_\alpha{}^{\dot\beta} [ \omega^\alpha{}_\gamma
\Phi^{\alpha(2)\gamma\dot\alpha(2)} + 
\omega^{\dot\alpha}{}_{\dot\gamma} 
\Phi^{\alpha(3)\dot\alpha\dot\gamma}] \nonumber \\
 && + 3\Lambda [ 2 \Phi_{\alpha\gamma(2)\dot\alpha(2)} 
e^\alpha{}_{\dot\beta} h_\beta{}^{\dot\beta} 
\Phi^{\beta\gamma(2)\dot\alpha(2)} - 
\Phi_{\alpha(3)\dot\alpha\dot\gamma} e_\beta{}^{\dot\alpha}
h^\beta{}_{\dot\beta} \Phi^{\alpha(3)\dot\beta\dot\gamma}] + h.c.
\end{eqnarray}
Here the first line contains non-minimal interactions (where graviton
enters only by curvatures $R_{\alpha(2)}$), while two remaining lines
exactly correspond to the standard substitution rules. The result
obtained is singular in the flat limit $\Lambda \to 0$, but rescaling
a coupling constant (which was previously set to 1) one can obtain the
vertex containing non-minimal term only\footnote{See \cite{KhZ20a} on
the flat limit for three arbitrary spins.}:
\begin{equation}
{\cal L}_1 \sim R_{\alpha\beta} \Phi^{\alpha\dot\alpha(4)}
\Phi^\beta{}_{\dot\alpha(4)} + h.c.. 
\end{equation}

\subsection{Second partially massless case}

As it was already mentioned in the introduction, applying the
Fradkin-Vasiliev formalism to massive or partially massless fields
leads to a lot of ambiguities related to field redefinitions
containing Stueckelberg zero-forms \cite{Zin24a}. In this
subsection we use a fermionic analogue \cite{KhZ19} of the so-called
Skvortsov-Vasiliev formalism \cite{SV06}, where all Stueckelberg
fields are absent, and the Fradkin-Vasiliev formalism works as
in the massless case. \\
{\bf Deformations for spin 7/2} also correspond to the standard
substitution rules:
\begin{eqnarray}
\Delta {\cal F}^{\alpha(4)\dot\alpha} &=& \omega^\alpha{}_\beta
\Phi^{\alpha(3)\beta\dot\alpha} + \omega^{\dot\alpha}{}_{\dot\beta}
\Phi^{\alpha(4)\dot\beta} - \frac{3\Lambda}{8} h^\alpha{}_{\dot\beta}
\Phi^{\alpha(3)\dot\alpha\dot\beta} + \frac{3c_3}{40}
h^{\alpha\dot\alpha}\Phi^{\alpha(3)}, \nonumber \\
\Delta {\cal F}^{\alpha(3)\dot\alpha(2)} &=& \omega^\alpha{}_\beta
\Phi^{\alpha(2)\beta\dot\alpha(2)} + \omega^{\dot\alpha}{}_{\dot\beta}
\Phi^{\alpha(3)\dot\alpha\dot\beta} + h_\beta{}^{\dot\alpha}
\Phi^{\alpha(3)\beta\dot\alpha} + \frac{M}{3} h^\alpha{}_{\dot\beta} 
\Phi^{\alpha(2)\dot\alpha(2)\dot\beta} + \frac{c_3}{8}
h^{\alpha\dot\alpha} \Phi^{\alpha(2)\dot\alpha}, \nonumber \\
\Delta {\cal F}^{\alpha(3)} &=& \omega^\alpha{}_\beta 
\Phi^{\alpha(2)\beta} - \frac{15\Lambda}{8} h^\alpha{}_{\dot\alpha}
\Phi^{\alpha(2)\dot\alpha} + 3c_3 h_{\beta\dot\alpha}
\Phi^{\alpha(3)\beta\dot\alpha}, \\
\Delta {\cal F}^{\alpha(2)\dot\alpha} &=& \omega^\alpha{}_\beta
\Phi^{\alpha\beta\dot\alpha} + \omega^{\dot\alpha}{}_{\dot\beta}
\Phi^{\alpha(2)\dot\beta} + h_\beta{}^{\dot\alpha}
\Phi^{\alpha(2)\beta} + \frac{2M}{3} h^\alpha{}_{\dot\beta}
\Phi^{\alpha\dot\alpha\dot\beta} + c_3 h_{\beta\dot\beta}
\Phi^{\alpha(2)\beta\dot\alpha\dot\beta}. \nonumber 
\end{eqnarray}
{\bf Deformations for graviton} are also unique up to one arbitrary
coupling constant
\begin{eqnarray}
\Delta R^{\alpha(2)} &=& a_0 
[\Phi^{\alpha\beta(3)\dot\alpha} \Phi^\alpha{}_{\beta(3)\dot\alpha} -
\frac{9\Lambda}{16} \Phi^{\alpha\beta(2)\dot\alpha(2)} 
\Phi^\alpha{}_{\beta(2)\dot\alpha(2)} - \frac{3}{40}
\Phi^{\alpha\beta(2)} \Phi^\alpha{}_{\beta(2)} + \frac{9\Lambda}{32}
\Phi^{\alpha\beta\dot\alpha} \Phi^\alpha{}_{\beta\dot\alpha} \nonumber
\\
 && \quad + \frac{9\Lambda}{64} \Phi^{\alpha\dot\alpha(2)} 
\Phi^\alpha{}_{\dot\alpha(2)} - \frac{3\Lambda}{8}
\Phi^{\alpha\beta\dot\alpha(3)} \Phi^\alpha{}_{\beta\dot\alpha(3)}
 + \frac{1}{4} \Phi^{\alpha\dot\alpha(4)} 
\Phi^\alpha{}_{\dot\alpha(4)}  ], \nonumber \\
\Delta T^{\alpha\dot\alpha} &=& a_0 [ - \frac{3}{8}
\Phi^{\alpha\beta(3)\dot\beta}
\Phi_{\beta(3)\dot\beta}{}^{\dot\alpha}  + \frac{3c_3}{40\Lambda} 
\Phi^{\alpha\beta(3)\dot\alpha} \Phi_{\beta(3)} 
 - \frac{3M}{16} \Phi^{\alpha\beta(2)\dot\beta(2)}
\Phi_{\beta(2)\dot\beta(2)}{}^{\dot\alpha} \\
 && + \frac{9c_3}{64} \Phi^{\alpha\beta(2)\dot\alpha\dot\beta}
\Phi_{\beta(2)\dot\beta}
 + \frac{9}{64} \Phi^{\alpha\beta(2)} \Phi_{\beta(2)}{}^{\dot\alpha} +
\frac{3M}{16} \Phi^{\alpha\beta\dot\beta} 
\Phi_{\beta\dot\beta}{}^{\dot\alpha} + h.c.]. \nonumber
\end{eqnarray}
Now we consider the deformed Lagrangian
\begin{eqnarray}
\hat{\cal L}_0 &=& - \frac{2}{3M\Lambda} 
\hat{\cal F}_{\alpha(4)\dot\alpha} \hat{\cal F}^{\alpha(4)\dot\alpha}
+ \frac{1}{2M} \hat{\cal F}_{\alpha(3)\dot\alpha(2)} 
\hat{\cal F}^{\alpha(3)\dot\alpha(2)} + \frac{1}{15M\Lambda} 
\hat{\cal F}_{\alpha(3)} \hat{\cal F}^{\alpha(3)} \nonumber \\
 && - \frac{3}{8M} \hat{\cal F}_{\alpha(2)\dot\alpha} 
\hat{\cal F}^{\alpha(2)\dot\alpha} - \frac{1}{4\Lambda} 
\hat{R}_{\alpha(2)} \hat{R}^{\alpha(2)} + h.c.
\end{eqnarray}
and require it to be gauge invariant. Non vanishing on-shell
variations are
\begin{equation}
\delta \hat{\cal F}^{\alpha(4)\dot\alpha} = R^\alpha{}_\beta
\zeta^{\alpha(3)\beta\dot\alpha}, \qquad
\delta \hat{R}^{\alpha(2)} = 2a_0 {\cal F}^{\alpha\beta(3)\dot\alpha}
\zeta^\alpha{}_{\beta(3)\dot\alpha}.
\end{equation}
They produce
\begin{equation}
\delta \hat{\cal L}_0 = [ \frac{16}{3M\Lambda} - \frac{2a_0}{\Lambda}]
R_{\alpha\beta} {\cal F}^{\alpha\gamma(3)\dot\alpha}
\zeta^\beta{}_{\gamma(3)\dot\alpha},
\end{equation}
so we put
$$
a_0 = \frac{8}{3M}. 
$$
At last, we extract a cubic part of the deformed Lagrangian and obtain
\begin{eqnarray}
M{\cal L}_1 &=& - \frac{4}{3\Lambda} {\cal F}_{\alpha(4)\dot\alpha}
[ \omega^\alpha{}_\beta \Phi^{\alpha(3)\beta\dot\alpha} +
\omega^{\dot\alpha}{}_{\dot\beta} \Phi^{\alpha(4)\dot\beta} -
\frac{3\Lambda}{8} h^\alpha{}_{\dot\beta}
\Phi^{\alpha(3)\dot\alpha\dot\beta} + \frac{3c_3}{40}
h^{\alpha\dot\alpha}\Phi^{\alpha(3)} ] \nonumber  \\
 && + {\cal F}_{\alpha(3)\dot\alpha(2)} [\omega^\alpha{}_\beta
\Phi^{\alpha(2)\beta\dot\alpha(2)} + \omega^{\dot\alpha}{}_{\dot\beta}
\Phi^{\alpha(3)\dot\alpha\dot\beta} + h_\beta{}^{\dot\alpha}
\Phi^{\alpha(3)\beta\dot\alpha} + \frac{M}{3} h^\alpha{}_{\dot\beta} 
\Phi^{\alpha(2)\dot\alpha(2)\dot\beta} + \frac{c_3}{8}
h^{\alpha\dot\alpha} \Phi^{\alpha(2)\dot\alpha} ] \nonumber  \\
 && + \frac{2}{15\Lambda} {\cal F}_{\alpha(3)} [ \omega^\alpha{}_\beta
\Phi^{\alpha(2)\beta} - \frac{15\Lambda}{8} h^\alpha{}_{\dot\alpha}
\Phi^{\alpha(2)\dot\alpha} + 3c_3 h_{\beta\dot\alpha}
\Phi^{\alpha(3)\beta\dot\alpha} ] \nonumber  \\
 && - \frac{3}{4} {\cal F}_{\alpha(2)\dot\alpha} [
\omega^\alpha{}_\beta \Phi^{\alpha\beta\dot\alpha} +
\omega^{\dot\alpha}{}_{\dot\beta} \Phi^{\alpha(2)\dot\beta} +
h_\beta{}^{\dot\alpha} \Phi^{\alpha(2)\beta} + \frac{2M}{3}
h^\alpha{}_{\dot\beta} \Phi^{\alpha\dot\alpha\dot\beta} + c_3
h_{\beta\dot\beta} \Phi^{\alpha(2)\beta\dot\alpha\dot\beta} \nonumber
\\
 && - \frac{4}{3\Lambda} R_{\alpha(2)} [
\Phi^{\alpha\beta(3)\dot\alpha}
\Phi^\alpha{}_{\beta(3)\dot\alpha} - \frac{9\Lambda}{16}
\Phi^{\alpha\beta(2)\dot\alpha(2)} 
\Phi^\alpha{}_{\beta(2)\dot\alpha(2)} - \frac{3}{40}
\Phi^{\alpha\beta(2)} \Phi^\alpha{}_{\beta(2)} + \frac{9\Lambda}{32}
\Phi^{\alpha\beta\dot\alpha} \Phi^\alpha{}_{\beta\dot\alpha} \nonumber
\\
 && \qquad \qquad + \frac{9\Lambda}{64} \Phi^{\alpha\dot\alpha(2)} 
\Phi^\alpha{}_{\dot\alpha(2)} - \frac{3\Lambda}{8}
\Phi^{\alpha\beta\dot\alpha(3)} \Phi^\alpha{}_{\beta\dot\alpha(3)}
 + \frac{1}{4} \Phi^{\alpha\dot\alpha(4)} 
\Phi^\alpha{}_{\dot\alpha(4)}  ]. 
\end{eqnarray}
Note that, contrary to the massless case, this vertex contains terms
with no more than four derivatives. Applying the same technique as in
the massless case, this vertex can be reduced to the sum of
non-minimal terms and all the terms corresponding to the standard
substitution rules. The resulting non-minimal terms look like:
\begin{equation}
{\cal L}_{non-min} = 2 R_{\alpha\beta} [ - \frac{2}{3\Lambda}
\Phi^{\alpha\dot\alpha(4)} \Phi^\beta{}_{\dot\alpha(4)} +
\Phi^{\alpha\gamma\dot\alpha(3)}
\Phi^\beta{}_{\gamma\dot\alpha(3)} - \frac{3}{8}
\Phi^{\alpha\dot\alpha(2)} \Phi^\beta{}_{\dot\alpha(2)}] + h.c.
\end{equation}
By construction the vertex is invariant under all gauge
transformations except $\zeta^{\alpha(5)}$ which produces variations
\begin{equation}
\delta {\cal L}_1  = - \frac{8}{3\Lambda} R_{\alpha\beta}
e^\beta{}_{\dot\alpha} \Phi^{\beta(4)\dot\alpha} 
\zeta^\alpha{}_{\beta(4)}, 
\end{equation}
which can be compensated by the corrections
\begin{equation}
\delta h^{\alpha\dot\alpha} \sim \Phi^{\beta(4)\dot\alpha}
\zeta^\alpha{}_{\beta(4)} + h.c. 
\end{equation}

\subsection{First partially massless case}

In this case we also use the Skvortsov-Vasiliev approach. \\
{\bf Deformations for spin 7/2} are again minimal:
\begin{eqnarray}
\Delta {\cal F}^{\alpha(3)\dot\alpha(2)} &=& \omega^\alpha{}_\beta
\Phi^{\alpha(2)\beta\dot\alpha(2)} + \omega^{\dot\alpha}{}_{\dot\beta}
\Phi^{\alpha(3)\dot\alpha\dot\beta} + \frac{M}{3} 
h^\alpha{}_{\dot\beta} \Phi^{\alpha(2)\dot\alpha(2)\dot\beta} + 
\frac{c_3}{8} h^{\alpha\dot\alpha} \Phi^{\alpha(2)\dot\alpha},
\nonumber \\
\Delta {\cal F}^{\alpha(2)\dot\alpha} &=& \omega^\alpha{}_\beta
\Phi^{\alpha\beta\dot\alpha} + \omega^{\dot\alpha}{}_{\dot\beta}
\Phi^{\alpha(2)\dot\beta} + c_3 h_{\beta\dot\beta}
\Phi^{\alpha(2)\beta\dot\alpha\dot\beta} + \frac{2M}{3} 
h^\alpha{}_{\dot\beta} \Phi^{\alpha\dot\alpha\dot\beta} + 
\frac{c_2}{3} h^{\alpha\dot\alpha} \Phi^\alpha, \\
\Delta {\cal F}^\alpha &=& \omega^\alpha{}_\beta \Phi^\beta + c_2
h_{\beta\dot\alpha} \Phi^{\alpha\beta\dot\alpha} + 2M
h^\alpha{}_{\dot\alpha} \Phi^{\dot\alpha}. \nonumber
\end{eqnarray}
{\bf Deformations for graviton} contain one arbitrary coupling
constant:
\begin{eqnarray}
\Delta R^{\alpha(2)} &=& a_0 [\Phi^{\alpha\beta(2)\dot\alpha(2)}
\Phi^\alpha{}_{\beta(2)\dot\alpha(2)} - \frac{1}{2}
\Phi^{\alpha\beta\dot\alpha} \Phi^\alpha{}_{\beta\dot\alpha} +
\frac{1}{6} \Phi^\alpha \Phi^\alpha \nonumber \\
 && \quad - \frac{1}{4}\Phi^{\alpha\dot\alpha(2)} 
\Phi^\alpha{}_{\dot\alpha(2)} + \frac{2}{3}
\Phi^{\alpha\beta\dot\alpha(3)} \Phi^\alpha{}_{\beta\dot\alpha(3)}],
 \\
\Delta T^{\alpha\dot\alpha} &=& \frac{a_0}{\Lambda} [\frac{M}{3}
\Phi^{\alpha\beta(2)\dot\beta(2)}
\Phi_{\beta(2)\dot\beta(2)}{}^{\dot\alpha} - \frac{M}{3}
\Phi^{\alpha\beta\dot\beta} \Phi_{\beta\dot\beta}{}^{\dot\alpha} + 
\frac{M}{3} \Phi^\alpha \Phi^{\dot\alpha} \nonumber \\
 && \quad + \frac{c_3}{4} (\Phi^{\alpha\beta(2)\dot\alpha\dot\beta} 
\Phi_{\beta(2)\dot\beta} + \Phi^{\alpha\beta\dot\alpha\dot\beta(2)}
\Phi_{\beta\dot\beta(2)}) - \frac{c_2}{6}
(\Phi^{\alpha\beta\dot\alpha} \Phi_\beta +
\Phi^{\alpha\dot\alpha\dot\beta} \Phi_{\dot\beta}) ]. \nonumber
\end{eqnarray}
Then we consider the deformed Lagrangian
\begin{equation}
\hat{\cal L}_0 = \frac{1}{2M} {\cal F}_{\alpha(3)\dot\alpha(2)}
{\cal F}^{\alpha(3)\dot\alpha(2)} - \frac{3}{8M} 
{\cal F}_{\alpha(2)\dot\alpha} {\cal F}^{\alpha(2)\dot\alpha} 
+ \frac{1}{4M} {\cal F}_\alpha {\cal F}^\alpha - \frac{1}{4\Lambda}
\hat{R}_{\alpha(2)} \hat{R}^{\alpha(2)} + h.c.
\end{equation}
and require it to be gauge invariant. Non vanishing on-shell
variations
\begin{equation}
\delta \hat{\cal F}^{\alpha(3)\dot\alpha(2)} = R^\alpha{}_\beta
\zeta^{\alpha(2)\beta\dot\alpha(2)}, \qquad
\delta \hat{R}^{\alpha(2)} = 2a_0 
{\cal F}^{\alpha\beta(2)\dot\alpha(2)} 
\zeta^\alpha{}_{\beta(2)\dot\alpha(2)}.
\end{equation}
They produce
\begin{equation}
\delta \hat{\cal L}_0 = - [\frac{3}{M} + \frac{2a_0}{\Lambda}]
R_{\alpha\beta} {\cal F}^{\alpha\gamma(2)\dot\alpha(2)}
\zeta^\beta{}_{\gamma(2)\dot\alpha(2)}, 
\end{equation}
SO we put
$$
a_0 = - \frac{3\Lambda}{2M}.
$$
A cubic part of the deformed Lagrangian has the form:
\begin{eqnarray}
M {\cal L}_1 &=& {\cal F}_{\alpha(3)\dot\alpha(2)}
[ \omega^\alpha{}_\beta \Phi^{\alpha(2)\beta\dot\alpha(2)}
+ \omega^{\dot\alpha}{}_{\dot\beta} 
\Phi^{\alpha(3)\dot\alpha\dot\beta} + \frac{M}{3} 
h^\alpha{}_{\dot\beta} \Phi^{\alpha(2)\dot\alpha(2)\dot\beta}
+ \frac{c_3}{8} h^{\alpha\dot\alpha} \Phi^{\alpha(2)\dot\alpha}]
\nonumber \\
 && - \frac{3}{4} {\cal F}_{\alpha(2)\dot\alpha}
 [ \omega^\alpha{}_\beta \Phi^{\alpha\beta\dot\alpha} +
\omega^{\dot\alpha}{}_{\dot\beta} \Phi^{\alpha(2)\dot\beta}
+ c_3 h_{\beta\dot\beta} \Phi^{\alpha(2)\beta\dot\alpha\dot\beta}
 + \frac{2M}{3} h^\alpha{}_{\dot\beta}
\Phi^{\alpha\dot\alpha\dot\beta} + \frac{c_2}{3} h^{\alpha\dot\alpha}
\Phi^\alpha ] \nonumber   \\
 && + \frac{1}{2} {\cal F}_\alpha [ \omega^\alpha{}_\beta \Phi^\beta
 + c_2 h_{\beta\dot\alpha} \Phi^{\alpha\beta\dot\alpha} + 2M
h^\alpha{}_{\dot\alpha} \Phi^{\dot\alpha}] \nonumber \\
 && + \frac{3}{4} R_{\alpha(2)}
[ \Phi^{\alpha\beta(2)\dot\alpha(2)}
\Phi^\alpha{}_{\beta(2)\dot\alpha(2)} - \frac{1}{2}
\Phi^{\alpha\beta\dot\alpha} \Phi^\alpha{}_{\beta\dot\alpha} +
\frac{1}{6} \Phi^\alpha \Phi^\alpha \nonumber \\
 && \qquad \qquad - \frac{1}{4}
\Phi^{\alpha\dot\alpha(2)} \Phi^\alpha{}_{\dot\alpha(2)} + \frac{2}{3}
\Phi^{\alpha\beta\dot\alpha(3)} \Phi^\alpha{}_{\beta\dot\alpha(3)} ].
\end{eqnarray}
Similarly to the previous cases, this vertex can be reduced to the sum
of non-minimal terms and terms corresponding to the standard
substitution rules. The resulting non-minimal terms look like:
\begin{equation}
{\cal L}_{non-min} = \frac{2}{M} R_{\alpha\beta} 
[ \Phi^{\alpha\gamma\dot\alpha(3)} \Phi^\beta{}_{\gamma\dot\alpha(3)}
- \frac{3}{8} \Phi^{\alpha\dot\alpha(2)} \Phi^\beta{}_{\dot\alpha(2)}]
+ h.c. 
\end{equation}
We see that this Lagrangian contains terms with at most two
derivatives, while in the previous cases the four derivative terms
were necessary. The reason for this is that by gauge fixing we
excluded extra fields, corresponding to higher derivative of the
physical fields. Now recall that by construction the vertex is
invariant under $\zeta^{\alpha(3)\dot\alpha(2)}$, 
$\zeta^{\alpha(2)\dot\alpha}$ and $\zeta^\alpha$ transformations but
not under $\zeta^{\alpha(4)\dot\alpha}$ and $\zeta^{\alpha(3)}$ 
transformations. Without these invariances, we cannot guarantee the
correct number of physical degrees of freedom. We will see the correct
result for this partially massless case in the following subsection.

\subsection{Massive case}

In the massive case, all gauge symmetries are spontaneously broken
and there is no analogue of the Skvortsov-Vasiliev formalism. In order
to resolve the ambiguities associated with field redefinitions, we
restrict ourselves to the minimal vertex (i.e. vertex with the minimum
number of derivatives possible), and follow a  down-up approach. We
start with the Lagrangian (\ref{lag}) and gauge transformations
(\ref{gauge}) and curvatures $(\ref{cur})$ where both the frame 
$e^{\alpha\dot\alpha}$ and the Lorentz covariant derivative $D$ are
dynamical. We still assume torsion to be zero, so that the only
source of non-invariance of the Lagrangian is the non-commutativity of
covariant derivatives. By straightforward calculation, we
obtain the following variation of the minimal Lagrangian
\begin{eqnarray}
\delta \hat{\cal L}_0 &=&  - 2e_\beta{}^{\dot\beta} 
[ R_\alpha{}^\gamma \Phi_{\alpha\gamma\dot\alpha(2)\dot\beta} + 
R_{\dot\alpha}{}^{\dot\gamma} 
\Phi_{\alpha(2)\dot\alpha\dot\beta\dot\gamma} +
R_{\dot\beta}{}^{\dot\gamma}
\Phi_{\alpha(2)\dot\alpha(2)\dot\gamma}]
\zeta^{\alpha(2)\beta\dot\alpha(2)} \nonumber \\
 && + 2  e_\beta{}^{\dot\beta} ( R_\alpha{}^\gamma
\Phi_{\gamma\dot\alpha\dot\beta} + R_{\dot\alpha}{}^{\dot\gamma}
\Phi_{\alpha\dot\beta\dot\gamma} + R_{\dot\beta}{}^{\dot\gamma}
\Phi_{\alpha\dot\alpha\dot\gamma}] \zeta^{\alpha\beta\dot\alpha} 
- 2 e_\alpha{}^{\dot\alpha} R_{\dot\alpha}{}^{\dot\beta}
\Phi_{\dot\beta} \zeta^\alpha. 
\end{eqnarray}
Now we introduce the following ansatz for non-minimal terms
\begin{equation}
{\cal L}_1 = R_{\alpha\beta} [ \kappa_1 
\Phi^{\alpha\dot\alpha(4)} \Phi^\beta{}_{\dot\alpha(4)} +
\kappa_2 \Phi^{\alpha\gamma\dot\alpha(3)}
\Phi^\beta{}_{\gamma\dot\alpha(3)} + \kappa_3
\Phi^{\alpha\dot\alpha(2)} \Phi^\beta{}_{\dot\alpha(2)}] + h.c.
\end{equation}
By direct calculations we found that all variations 
$\delta (\hat{\cal L}_0 + {\cal L}_1)$ can be compensated by
corrections to graviton transformations provided
\begin{equation}
\kappa_1 = \frac{120}{Mc_0{}^2}, \qquad
\kappa_2 = \frac{2}{M}, \qquad
\kappa_3 = - \frac{3}{4M}.
\end{equation}
In this, the required corrections correspond to the following torsion
deformations:
\begin{eqnarray}
\Delta T^{\alpha\dot\alpha} &=& \frac{1}{2} [ - \kappa_1
\Phi^{\alpha\beta(4)} \Phi_{\beta(4)}{}^{\dot\alpha} - \frac{2}{M} 
\Phi^{\alpha\beta(3)\dot\beta} \Phi_{\beta(3)\dot\beta}{}^{\dot\alpha}
 - \Phi^{\alpha\beta(2)\dot\beta(2)}
\Phi_{\beta(2)\dot\beta(2)}{}^{\dot\alpha} \nonumber \\
 && - \frac{3c_3}{10}\kappa_1 \Phi^{\alpha\beta(3)\dot\alpha}
\Phi_{\beta(3)} + \frac{3c_3}{4M} 
\Phi^{\alpha\beta(2)\dot\alpha\dot\beta} \Phi_{\beta(2)\dot\beta}
\nonumber \\
 && - \frac{3}{4M} \Phi^{\alpha\beta(2)} 
\Phi_{\beta(2)}{}^{\dot\alpha}
+ \Phi^{\alpha\beta\dot\beta} \Phi_{\beta\dot\beta}{}^{\dot\alpha} 
 + \frac{c_2}{2M} \Phi^{\alpha\beta\dot\alpha} \Phi_\beta
 - \Phi^\alpha \Phi^{\dot\alpha} ]. 
\end{eqnarray}
Note that this result is completely consistent with the results for
massless and partially massless cases (where they overlap). Resulting
non-minimal terms look like
\begin{equation}
{\cal L}_1 = \frac{2}{\sqrt{m^2-9\Lambda}} R_{\alpha\beta} 
[ \frac{2}{(m^2-8\Lambda)} 
\Phi^{\alpha\dot\alpha(4)} \Phi^\beta{}_{\dot\alpha(4)} +
 \Phi^{\alpha\gamma\dot\alpha(3)}
\Phi^\beta{}_{\gamma\dot\alpha(3)} - \frac{3}{8}
\Phi^{\alpha\dot\alpha(2)} \Phi^\beta{}_{\dot\alpha(2)}] + h.c.
\end{equation}
First of all, we note the same general properties as for spin 5/2 and
spin 3. In anti de Sitter space $\Lambda < 0$ the vertex has a
non-singular massless limit, while for non-zero mass it has a
non-singular flat limit. There is a singularity corresponding to the
points at the boundary of a unitary forbidden region 
$m^2 = 9\Lambda$. In these case, the only way to obtain a
non-trivial result is to rescale the gravitational coupling
constant (which was set to  1 previously). This leaves only
non-minimal terms:
\begin{equation}
{\cal L}_1 \sim  R_{\alpha\beta} [ \frac{2}{\Lambda} 
\Phi^{\alpha\dot\alpha(4)} \Phi^\beta{}_{\dot\alpha(4)} +
 \Phi^{\alpha\gamma\dot\alpha(3)}
\Phi^\beta{}_{\gamma\dot\alpha(3)} - \frac{3}{8}
\Phi^{\alpha\dot\alpha(2)} \Phi^\beta{}_{\dot\alpha(2)}] + h.c.
\end{equation}
At last, one more singularity corresponds to the first partially
massless case $m^2 = 8\Lambda$. In this case the only non-trivial
vertex (obtained again by rescaling of the coupling constant) is
simply
\begin{equation}
{\cal L}_1 \sim R_{\alpha\beta} \Phi^{\alpha\dot\alpha(4)}
\Phi^\beta{}_{\dot\alpha(4)} + h.c.
\end{equation}
which drastically differs from the result of previous subsection.

\section{Conclusion}

In this work, we extended our previous results on gravitational
interactions for massive spin 5/2 and spin 3 to massive spin 7/2,
including all its massless and partially massless limits. As 
expected, the results obtained share a number of properties, such as a
non-singular massless limit in $AdS$, a non-singular flat limit for 
non-zero mass, and a singularity at the points corresponding to
the boundary of the unitary forbidden region. There exists also a
singularity at the points corresponding to the first partially
massless limit (inside the forbidden region). One reason for
considering spin 7/2 was that, in both previous cases, only the
highest helicities required non-minimal terms, so it was unclear how
the structure of non-minimal terms for an arbitrary spin would look
like. Based on the result on spin 7/2, we can suggest that the
structure must simply be the sum of the  necessary non-minimal terms
required for each massless helicity $5/2 \le h \le s$. If this is
true, then  generalizing  our results to arbitrary spins becomes a
purely combinatorical task which we leave as an exercise to the
reader.

One more subject of interest in this paper was the so-called
Skvortsov-Vasiliev formalism for describing  partially massless
fields. It does not include any Stueckelberg fields, so the
construction of interactions goes without ambiguities exactly as in
the massless case. Indeed, this formalism has proven useful
in our investigations. However, it arises form the general
formalism as a result of a partial gauge fixing when we set
Stueckelberg fields to zero and solve their equations. But to ensure
the correct number of physical degrees of freedom after constructing
an interaction, we still need to take care on symmetries which were
fixed. Our results in this paper show that this is not always
possible. The technical reason for this is that during the gauge
fixing, we exclude some the extra fields, corresponding to higher
derivatives of physical fields.

\end{document}